# Composition Driven Band Alignment Evolution in Grafted Si/GeSn Heterostructures

Pargam Vashishtha[1,#], Samuel A. Haessly[1,#], Yang Liu[1], Tsung-Han Tsai[1], Kushal Dahal[2], Shui-Qing Yu[2], Edward, K. Huang[3], Nanditha M. Dissanayake[3], Zhenqiang Ma[1,*]

[1]*Department of Electrical and Computer Engineering, University of Wisconsin-Madison, Madison, Wisconsin 53706, USA.*

[2]*Department of Electrical Engineering, University of Arkansas, Fayetteville, Arkansas 72701, USA*

[3]*Attollo Engineering, Camarillo, CA 93012, USA.*

[#]*Authors have contributed equally*

**Corresponding Author's Email: mazq@engr.wisc.edu*

## Abstract

Understanding and controlling band alignment in lattice-mismatched group-IV semiconductor-based heterostructures is fundamental to rational band-structure engineering, as it governs the electronic structure and carrier transport in heterointegrated semiconductor systems. Despite growing interest in Si/GeSn heterostructures, the composition-dependent band alignment has not been experimentally established. Here, we experimentally establish the evolution of band alignment in semiconductor-grafted Si/GeSn heterostructures with contents ranging from 8.8% to 12.4% using X-ray photoelectron spectroscopy and Kraut's core-level method. The valence-band offset increases systematically from ~0.29 to ~0.57 eV with increasing Sn composition, whereas the corresponding effective conduction-band offset decreases from ~0.30 to ~0.12 eV. The observed band-alignment evolution is attributed to an upward shift of the GeSn valence-band maximum arising from enhanced p-p orbital hybridization, reduced bonding-antibonding splitting, and compressive-strain-induced heavy-hole band lifting. These findings establish the experimental relationship between alloy composition and interfacial band alignment in semiconductor-grafted Si/GeSn heterostructures, providing a quantitative framework for rational band-structure engineering in lattice-mismatched group-IV semiconductor heterostructures.

## Introduction:

The continuous expansion of data traffic for optical interconnects, sensing, and emerging mid-infrared photonic technologies is driving silicon (Si) beyond its intrinsic material limits.[1, 2] In this context, group-IV alloy engineering has attracted considerable attention because it enables the introduction of optoelectronic functionality while preserving compatibility with the Si platform.[3] Among these materials, germanium-tin (GeSn) has emerged as a promising candidate owing to its composition and strain-dependent band structure.[4] With increasing Sn incorporation and appropriate strain engineering, GeSn can be transformed from an indirect- to a direct-bandgap semiconductor, enabling continuous bandgap tuning from the near-infrared to the short-wave and mid-infrared spectral regions.[5] This unique capability makes GeSn highly attractive for wavelength-extended photodetectors, light emitters, and high-speed electronic–photonic integrated devices.[6-8] For many heterostructure device architectures, especially avalanche electronic, optoelectronics and high-performance photonic components, the integration of GeSn with Si is essential because Si provides an excellent platform for carrier multiplication, low excess noise, and mature CMOS processing.[9] However, the epitaxial growth of GeSn directly on Si remains extremely challenging, mainly at high Sn compositions, due to the large lattice mismatch and the low solid solubility of Sn in Ge.[10] Conventional buffer-mediated approaches employ thick Ge layers to accommodate strain and suppress defect propagation, but these buffers introduce several critical limitations, including increased dark current, reduced carrier collection efficiency, and restricted avalanche gain.[11] Moreover, the target direct interface of GeSn and Si is not achievable, which constrains band-structure engineering and spectral tunability.

Semiconductor grafting has recently emerged as an alternative integration strategy for forming electronically active lattice-mismatched heterointerfaces.[12, 13] This approach enables the formation of high-quality interfaces without the need for thick strain-relaxation buffers and has been successfully applied to a wide range of dissimilar material systems.[14-20] By passivating surface dangling bonds and thereby reducing interface trap density, grafted heterostructures provide an ideal platform for experimentally probing interfacial band alignment and electronic structure.[21] For Si/GeSn heterojunctions, the interfacial band alignment critically determines carrier transport, impact ionization efficiency, optical absorption, and overall device performance. Although band offsets for grafted Si/GeSn have been reported for a fixed Sn composition in the alloy, a systematic understanding of the evolution of the valence-band alignment with varying Sn content is still lacking.[22] Moreover, the increased surface roughness at high Sn concentrations can lead to dielectric-layer discontinuities and interfacial damage during the grafting process. A comprehensive understanding of composition-dependent band offsets at chemically bonded grafted interfaces is therefore essential for the rational design of band-engineered heterostructures.

In this work, we investigate the evolution of band alignment in grafted Si/GeSn heterostructures with varying Sn compositions. High-quality GeSn epilayers with different Sn contents were chemically-mechanically polished to achieve smooth surfaces and subsequently were grafted with monocrystalline Si nanomembranes (NMs) through a deterministic transfer process. The interfacial electronic structure was probed by X-ray photoelectron spectroscopy, and the valence-band offsets were determined using Kraut's core-level method. This study reveals that increasing the Sn composition from 8.8% to 12.4% systematically increases the valence-band offset (VBO) from 0.23 to 0.57 eV, while the conduction-band offset (CBO) decreases from 0.30 eV to 0.12 eV. By directly correlating the extracted band offsets with the Sn composition and strain state of the GeSn layers, this work provides a comprehensive understanding of composition-dependent band alignment in grafted Si/GeSn heterostructures. It establishes a viable pathway for band-engineered group-IV optoelectronic devices. Unlike previous reports limited to a single alloy composition, this work experimentally establishes the evolution of band alignment across three technologically relevant GeSn compositions using a consistent grafting platform and identical XPS methodology.

## Experimental:

Epitaxial GeSn layers with varying Sn compositions were grown on chemically cleaned Si substrates by chemical vapor deposition (CVD) using an Epsilon® Advanced Epi system (Materials and Devices Ltd.). Before growth, the substrates were preheated, and a 1 μm Ge buffer layer was grown. Thereafter, 400-to-600 nm GeSn epitaxy was carried out. The structural quality (composition, strain/strain) of the as-grown films was evaluated by high-resolution X-ray diffraction (HRXRD) using a PANalytical X'Pert PRO MRD system with CuK$\alpha_1$ radiation ($\lambda$ = 1.5406 Å). The surface morphology was examined by atomic force microscopy (AFM, Bruker). Raman spectroscopy was performed using a Horiba LabRAM spectrometer with 633 nm excitation, and the optical properties were further investigated by photoluminescence (PL), was recorded at 300K under 532 nm laser excitation with a power of 200 mW, employing an InSb detector for signal acquisition.

To reduce the roughness of the as-grown samples, chemical-mechanical polishing (CMP) was performed. Polishing was carried out on a Kemet KemCol 15 CMP tool using a 380 mm Kemet HAF polishing pad and an in-house fixture designed for small samples. A 0.09 μm colloidal silica slurry (Col-K[NC]) was used with hydrogen peroxide added as an oxidizer (0.015% of slurry weight) to promote controlled chemical-mechanical removal and reduce surface tin-segregation features. CMP was then performed using an applied mass of ~482-488 gm, a platen speed of 30 rpm, and polishing durations ranging from 15 sec to 1 min. Slurry volumes of 20-25 mL were dispensed per run, and the polishing pad was pre-wetted with 100 mL of deionized water to promote uniform slurry distribution and stable polishing conditions.

The polished GeSn surfaces were used to fabricate Si/GeSn heterostructures via semiconductor grafting, following the procedure reported in our previous work.[22] The GeSn epi substrates were coated with $Al_2O_3$ for 5 cycles (~0.5 nm) by atomic layer deposition (ALD; Fiji G2, Cambridge Nanotech) @ 200°C . Then a 205 nm thick Si-NM was prepared from a silicon-on-insulator (SOI) wafer by selectively removing the buried $SiO_2$ layer using an HF-based wet etching process. The released Si-NM was subsequently transferred onto an $Al_2O_3$/GeSn substrate using a polydimethylsiloxane (PDMS) stamp. To enhance the interfacial adhesion and promote bonding, the heterostructures were subjected to rapid thermal annealing (RTA) at 250°C for 5 min. Optical microscopy images of the grafted heterostructures, acquired using an Olympus BX51M, are shown in Fig. S1, Supporting Information (SI).

X-ray photoelectron spectroscopy (XPS; Thermo Scientific K-Alpha) was employed to investigate the chemical composition and electronic structure of the heterointerfaces. Before analysis, chemically cleaned bare GeSn and bare Si-NM samples were cleaned in situ using a 2000 eV $Ar^+$ cluster ion beam for 15 Sec to remove native oxides and surface carbon contaminants. For interfacial measurements of the grafted heterostructures, the Si-NM overlayer was first thinned by reactive ion etching (RIE; Plasma-Therm 790) using a $CF_4/O_2$ plasma. This was followed by a controlled, in situ, cluster-mode $Ar^+$-ion etching process in the XPS chamber (2 mm raster size, 2 and 4 keV beam energies) to expose the buried interface. High-resolution core-level spectra of Ge 3d, Sn 3d, Sn 4d, and Si 2p, together with valence-band spectra, were collected; the same-intensity detection of both Si and Ge signals confirmed that the interface region had been reached. The binding energy scale was calibrated using the Au $4f_{7/2}$ peak at 84.0 eV. All spectra were acquired with a pass energy of 10 eV, an analysis area of 400 μm, an energy step size of 0.02 eV, and a maximum dwell time of 10 Sec. Data processing and peak deconvolution were performed using CasaXPS with Shirley background subtraction.

## Results and Discussion:

All as-grown samples were characterized by HRXRD to evaluate the Sn composition and strain state of the GeSn epilayers. The extracted Sn compositions of 10.2%, 11.7%, and 13.1% are hereafter denoted as *l*-GeSn, *m*-GeSn, and *h*-GeSn samples, respectively (Fig. S2, SI). AFM measurements were used to determine the root-mean-square surface roughness of the as-grown samples, as shown in the height profiles at a 10×10 $\mu m^2$ scale in Fig. 1(a-c). The roughness values of the as-grown *l*-GeSn, *m*-GeSn, and *h*-GeSn films were 3.93 nm, 9.24 nm, and 7.87 nm, respectively. Such surface roughness is not good enough for subsequent grating fabrication, as it can lead to discontinuities or pinching in the deposited dielectric layer. Therefore, a stepwise chemical-mechanical polishing (CMP) process was implemented for all samples, as schematically illustrated in Fig. 1(d). For polishing, the samples were mounted face-down in a customized wafer fixture, which was then placed on the CMP lapping tool. Polishing was carried out at 30 rpm for varying durations, with the slurry solution supplied periodically. Further experimental details are provided in the Experimental Section. The polishing was performed in a controlled stepwise manner to achieve a target surface roughness of ~1 nm without degrading the high Sn composition. The complete polishing parameters for each sample are summarised in Table S1, SI. After the final polishing process, roughness was reduced to 0.29 nm, 0.31 nm, and 1.56 nm for the *l*-GeSn, *m*-GeSn, and *h*-GeSn samples, respectively, as depicted in AFM height profile data in Fig. 1(e-g).

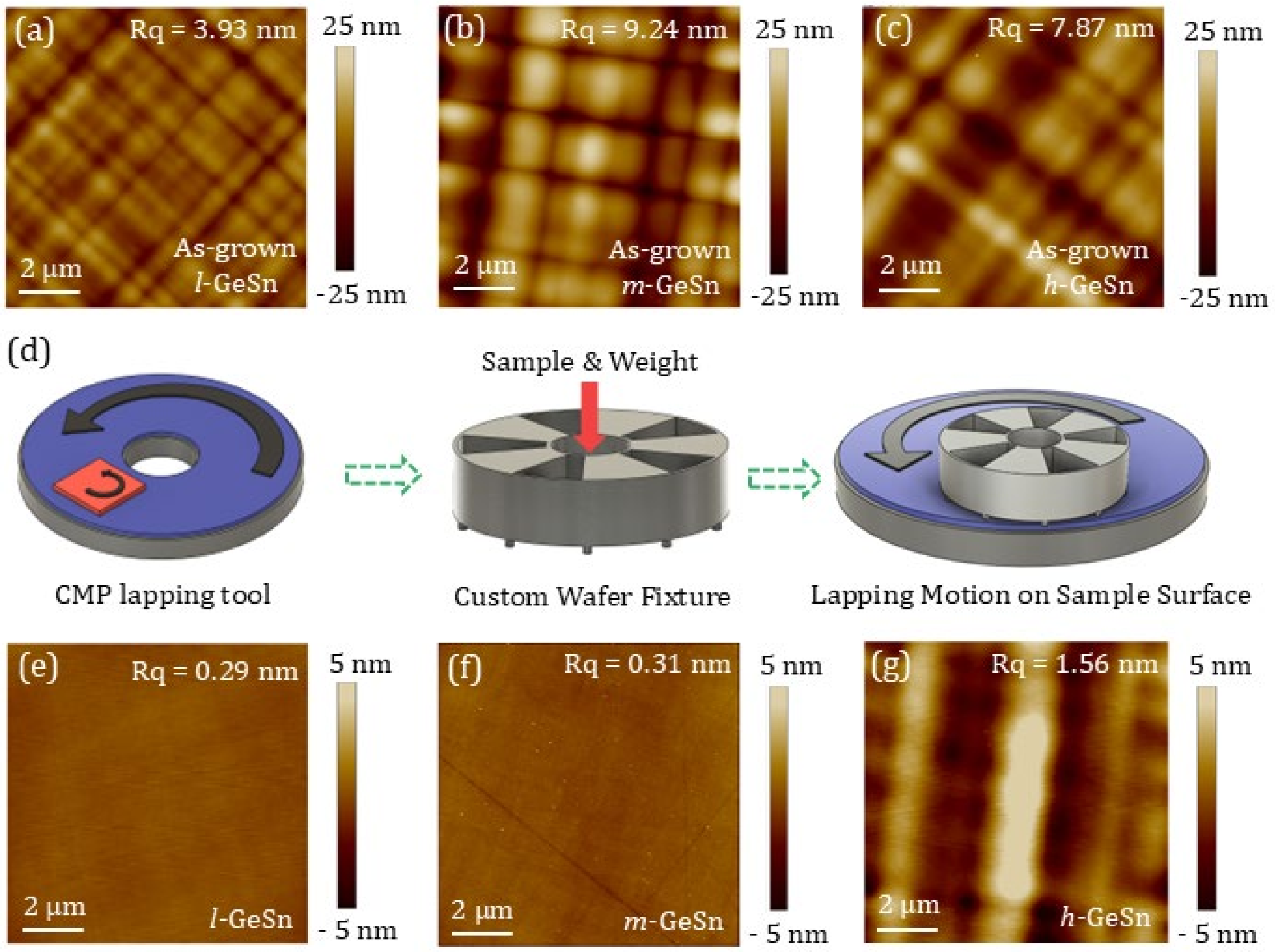

***Figure 1.*** *AFM height profile (10×10 µm²) images of as-grown GeSn thin films; (a) l-GeSn, (b) m-GeSn, and (c) h-GeSn. (d) Schematic of the chemical-mechanical polishing process. AFM height profile (10×10 µm²) polished samples (e) l-GeSn, (f) m-GeSn, and (g) h-GeSn.*

HRXRD measurements were performed on the polished samples, and the corresponding 2θ/ω around the (004) reflection are depicted in Fig. 2(a-c). In all samples, the sharp peak at ~69° originates from the Si (004) substrate, while the peak at ~66° corresponds to the Ge buffer layer. In addition to these features, distinct diffraction peaks from the GeSn epilayers are observed at lower 2θ values due to the larger lattice constant of GeSn compared to Ge. For the *l*-GeSn sample, a single intense GeSn peak is observed at 64.86°, as shown in Fig. 2(a), indicating good crystalline quality. Moreover, the *m*-GeSn (Fig. 2(b)) and *h*-GeSn (Fig. 2(c)) samples exhibit two and four GeSn-related diffraction peaks, respectively, revealing the presence of different Sn % layers in the sample, which lead to variations in the out-of-plane lattice parameter and hence different diffraction angles. Furthermore, reciprocal space mapping (RSM) around the asymmetric (-2-24) reflection was carried out to evaluate the composition and strain state of the GeSn thin films using Vegard's law (bowing parameter $-0.0166\times10^{-10}$), and the corresponding results are presented in Fig. 2(d–f). The analysis indicates that the polished samples exhibit surface Sn contents of 8.8%, 10.9%, and 12.4%, with corresponding compressive strain values of 0.34%, 0.26%, and 0.91% for the *l*-GeSn, *m*-GeSn, and *h*-GeSn layers, respectively. A detailed summary of the extracted structural and morphological parameters is provided in Table S2, SI. Raman spectroscopy was further employed to probe the vibrational characteristics of the polished GeSn thin films, as deconvoluted spectra are presented in Fig. 2(g-i). With increasing Sn incorporation, the Ge-Ge longitudinal optical phonon mode (~300 $cm^{-1}$) exhibits a reduction in intensity, reflecting lattice relaxation and alloy disorder induced by Sn incorporation.[23] In addition to this dominant feature, a broad spectral contribution emerges in the 200-280 $cm^{-1}$ region. The peak centered around ~260–280 $cm^{-1}$ is assigned to the Ge-Sn vibrational mode, while the feature near ~285 $cm^{-1}$ appears as a shoulder on the low-frequency side of the Ge-Ge peak, indicating the coexistence of Ge-Ge and Ge-Sn bonding environments.[24] Furthermore, a low-wavenumber mode around ~185 $cm^{-1}$ is observed, which is attributed to Sn-Sn vibrations.[24]

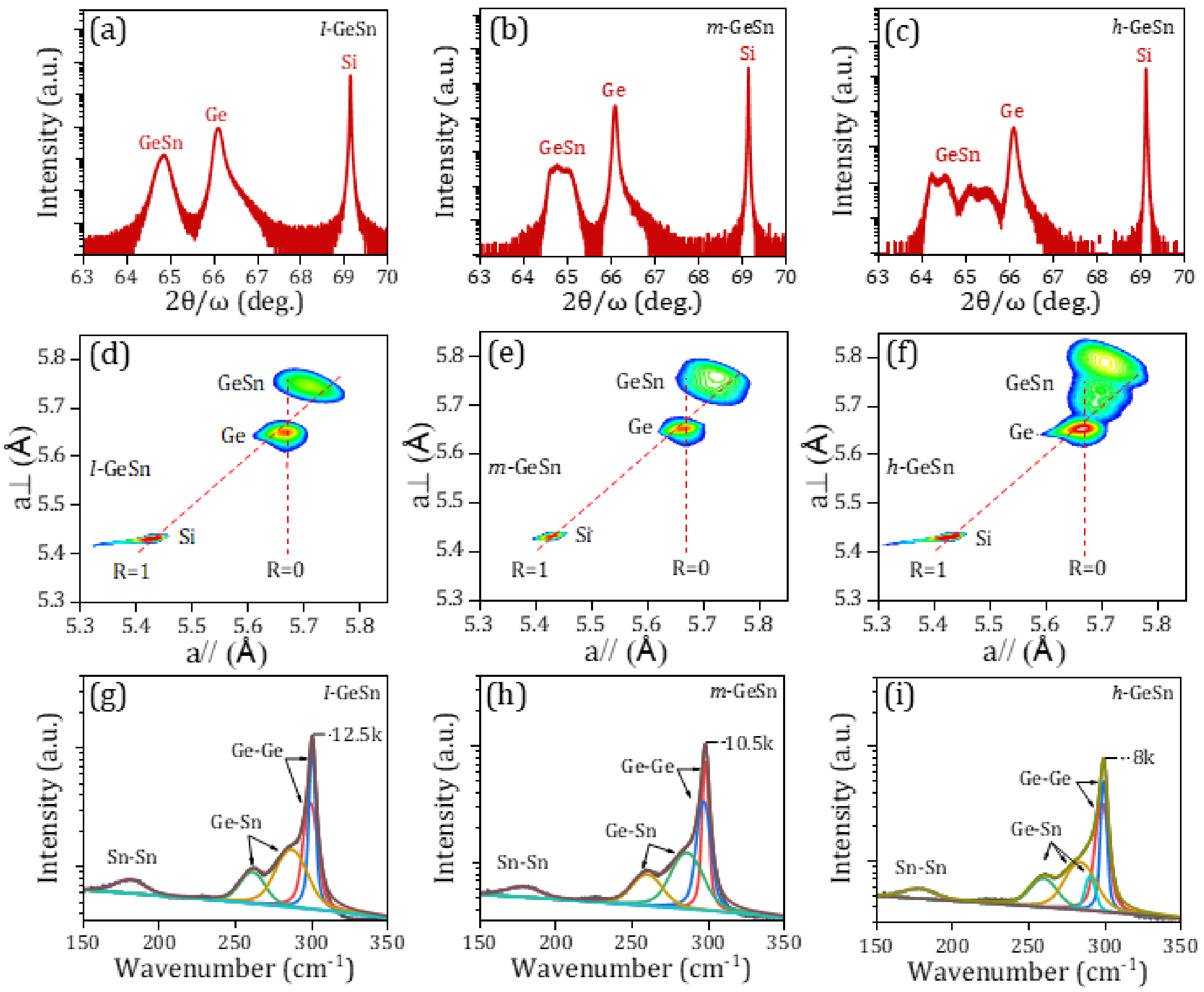

***Figure 2.*** *HRXRD 2θ/ω scans of the polished (a) l-GeSn, (b) m-GeSn, and (c) h-GeSn samples showing the Si, Ge, and GeSn (004) reflections. RSM around the asymmetric (-2-24) reflection for (d) l-GeSn, (e) m-GeSn, and (f) h-GeSn used to extract the Sn % and strain state. (g–i) Raman spectra of the samples exhibiting the Ge-Ge, Ge-Sn, and Sn-Sn vibrational modes. The numbers are counts.*

The above results confirm that the polished *l*-GeSn (8.8% Sn), *m*-GeSn (10.9% Sn), and *h*-GeSn (12.4% Sn) epilayers are relaxed and exhibit low surface roughness values of 0.29, 0.31, and 1.56 nm, respectively, suitable platforms for semiconductor grafting and reliable interfacial electronic characterization for the fabrication of grafted Si/GeSn heterostructures with varying Sn composition. The Si/$Al_2O_3$/GeSn heterostructures were fabricated via a multistep integration process (see Experimental section), as schematically illustrated in Fig. 3(a-f). This procedure involves the dry release of the undercut Si-NM using a PDMS stamp, followed by its deterministic transfer onto the $Al_2O_3$-coated surface to preserve electronic coupling while providing sufficient interface passivation, and onto the polished GeSn surface to form the two-sided-bonded grafted heterointerface. Moreover, before transfer, high-resolution XPS measurements were carried out on the bare Si-NM after in situ ion-gun treatment, followed by ex situ chemical cleaning. The deconvoluted Si 2p core-level spectrum (Fig. 3(g)) shows peak 2p $d_{3/2}$ at a binding energy of 99.49 eV, while the valence-band spectrum (Fig. 3(h)) yields a valence-band maximum (VBM) of 0.61 eV.

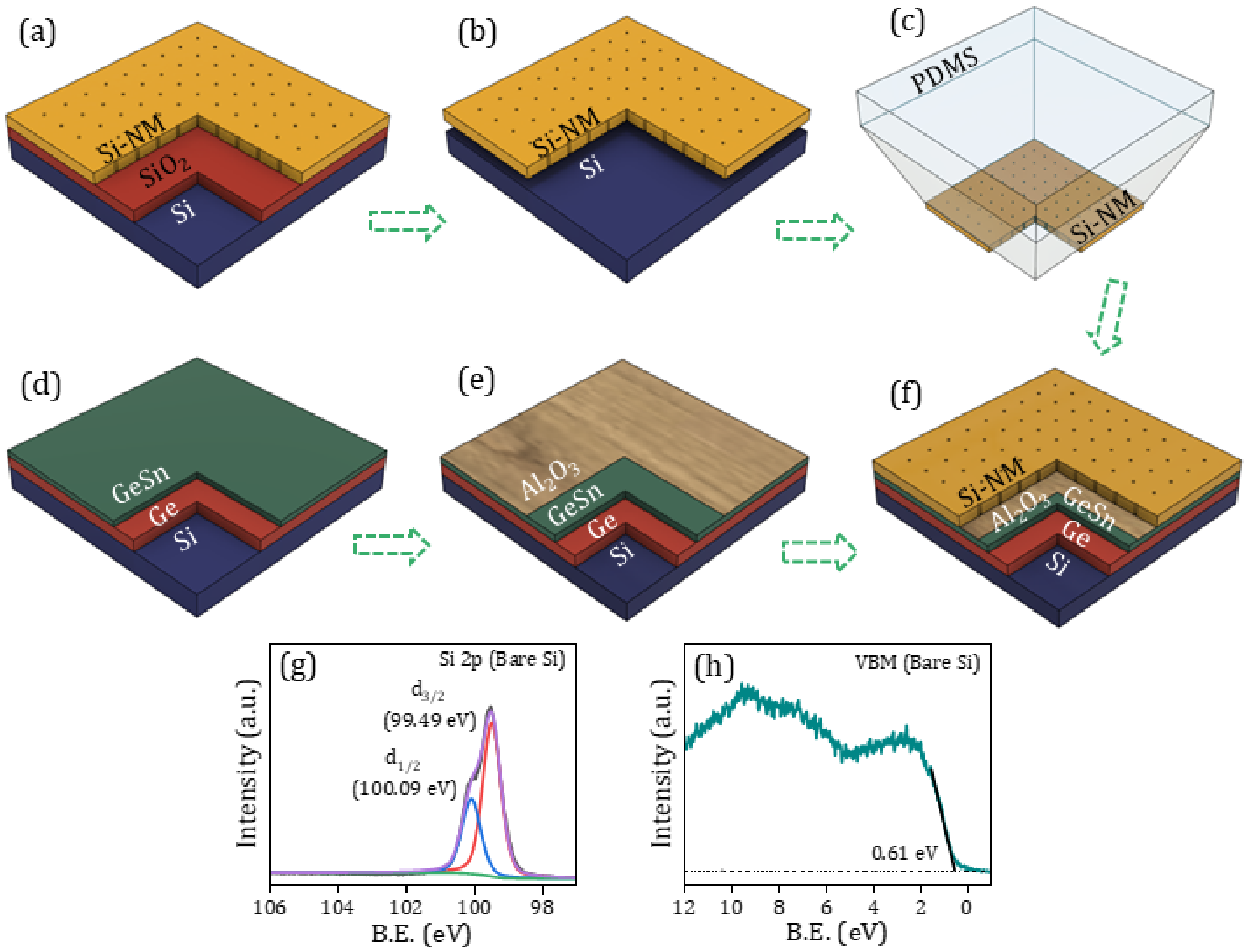

***Figure 3.*** *(a, b) Release of the undercut Si-NM from the SOI structure and its (c) pickup using a PDMS stamp. (d, e) deposition of* $Al_2O_3$ *on GeSn samples, and (f) transfer of the Si-NM to form the grafted Si/GeSn heterostructure. (g) High-resolution Si 2p core-level, and (h) Valence-band, spectrum of bare Si.*

High-resolution XPS measurements were performed on the cleaned, polished bare *l*-GeSn (8.8% Sn) sample. The deconvoluted spectra for the Ge 3d, Sn 3d, Sn 4d, and the valence-band region are presented in Fig. 4(a-d), respectively. The extracted binding energies are 29.31 eV for Ge $3d_{5/2}$, 485.09 eV for Sn $3d_{5/2}$, and 24.29 eV for Sn $4d_{5/2}$, while the VBM is located at 0.25 eV. Thereafter, the Si/*l*-GeSn heterostructure was fabricated using the grafting approach as described previously, and the interfacial electronic structure was examined by XPS depth profiling. Prior to the measurements, the Si-NM was thinned to ~188 nm by controlled RIE (see Experimental section). The sample was transferred to the XPS chamber in a nitrogen ambient to limit oxidation, followed by stepwise in situ depth profiling using a 4 keV $Ar^+$ cluster ion beam, as depicted in Fig. S4(a-d), SI. With increasing sputter depth, the Si 2p intensity decreased while the Ge 3d signal increased, evidencing the approach to the heterointerface. The interface region was operationally identified when comparable Si 2p and Ge 3d photoelectron intensities were simultaneously detected during depth profiling. The invariant binding energy and line shape of the Si 2p peak throughout the milling sequence indicate that the crystalline quality of the Si-NM was preserved during deep sputtering, confirming the reliability of the top-down XPS profiling method. After thinning, high-resolution core-level XPS spectra of Ge 3d, Sn 3d, Sn 4d, and Si 2p were recorded at the heterointerface using a fixed pass energy of 10 eV, as presented in Fig. 4(e-h). For the heterostructure, the Ge $3d_{5/2}$, Sn $3d_{5/2}$, and Sn $4d_{5/2}$ peaks are located at 29.68 eV (Fig. 4(e)), 485.45 eV (Fig. 4(f)), and 24.68 eV (Fig. 4(g)), respectively, while the Si $2p_{3/2}$ core-level peak appears at 99.79 eV (Fig. 4(h)). The shift in core-level binding energies for the heterostructure relative to its bare counterparts originates from charge redistribution at the interface.

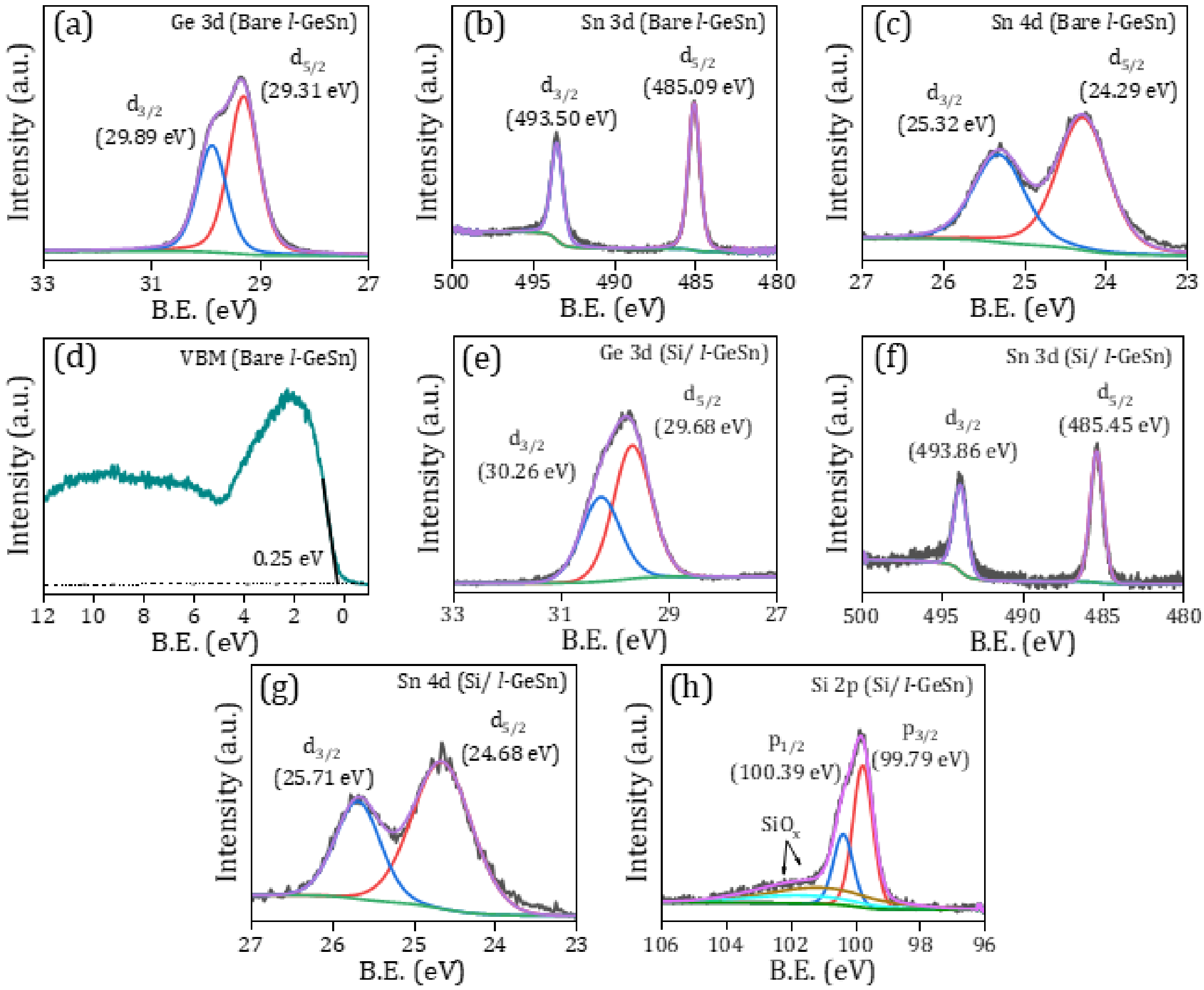

***Figure 4.*** *High-resolution XPS spectra of the bare l-GeSn surface showing (a) Ge 3d, (b) Sn 3d, (c) Sn 4d, and (d) the valence-band maxima. Corresponding core-level spectra acquired at the grafted Si/l-GeSn heterointerface: (e) Ge 3d, (f) Sn 3d, (g) Sn 4d, and (h) Si 2p.*

Similarly, high-resolution XPS measurements were performed on the cleaned and polished *m*-GeSn (10.9% Sn) epilayer. The Ge 3d, Sn 3d, Sn 4d, and valence-band spectra of the bare surface are shown in Fig. 5(a-d). The extracted binding energies are 29.41 eV for Ge $3d_{5/2}$, 485.19 eV for Sn $3d_{5/2}$, and 24.38 eV for Sn $4d_{5/2}$, with the VBM located at 0.22 eV. The core levels shift systematically toward higher binding energy, while the VBM moves to lower binding energy relative to bare *l*-GeSn, evidencing a composition-dependent modification of the electronic structure. The grafted Si/*m*-GeSn heterostructure was subsequently fabricated using the same deterministic grafting approach as previously. The interface was accessed by thinning the Si-NM to ~197 nm through controlled RIE, followed by in-situ sputtering with a 2/4 keV $Ar^+$ cluster ion beam. The corresponding depth profile (Fig. S5, SI) confirms the approach to the interfacial region. High-resolution interfacial spectra acquired at a pass energy of 10 eV (Fig. 5(e-h)) show the Ge $3d_{5/2}$, Sn $3d_{5/2}$, Sn $4d_{5/2}$, and Si $2p_{3/2}$ peaks at 29.70, 485.46, 24.69, and 99.78 eV, respectively. All core levels exhibit systematic shifts in binding energy toward higher values relative to their bare counterparts, consistent with charge redistribution at the interface.

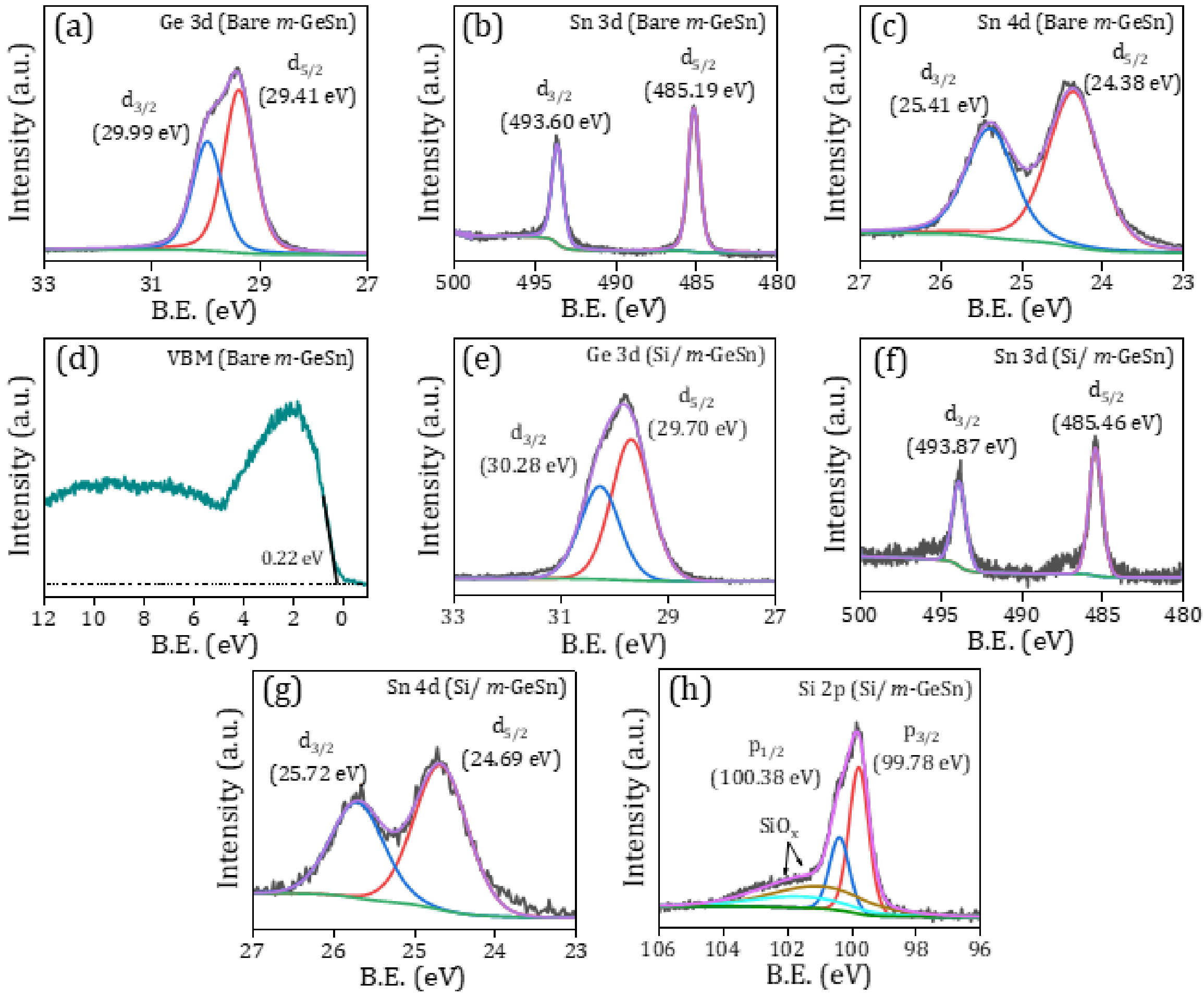

***Figure 5.*** *High-resolution XPS spectra of the bare m-GeSn surface showing (a) Ge 3d, (b) Sn 3d, (c) Sn 4d, and (d) the valence-band maxima. Corresponding core-level spectra acquired at the Si/m-GeSn heterointerface: (e) Ge 3d, (f) Sn 3d, (g) Sn 4d, and (h) Si 2p.*

Fig. 6(a-d) displays the high-resolution spectra of the polished bare *h*-GeSn (12.4 % Sn) surface. The Ge $3d_{5/2}$, Sn $3d_{5/2}$, and Sn $4d_{5/2}$ core levels are located at 29.57 eV, 485.30 eV, and 24.56 eV, respectively, while the VBM is positioned at 0.14 eV. Compared with the lower-Sn compositions, the core levels shift to higher binding energies, and the VBM moves further towards the Fermi level, indicating a progressive Sn-induced modification of the electronic structure. To probe the fabricated (using the same method as employed for other heterostructures) interface, the Si-NM was thinned by controlled RIE (~184 nm) followed by stepwise in-situ $Ar^+$ cluster-ion sputtering, and the approach to the interfacial region was confirmed by the depth profile (Fig. S6, SI) when the Si 2p and Ge 3d intensities became comparable. The interfacial spectra in Fig. 6(e-h) show the Ge $3d_{5/2}$, Sn $3d_{5/2}$, Sn $4d_{5/2}$, and Si $2p_{3/2}$ peaks at 29.77 eV, 485.48 eV, 24.73 eV, and 99.77 eV, respectively.

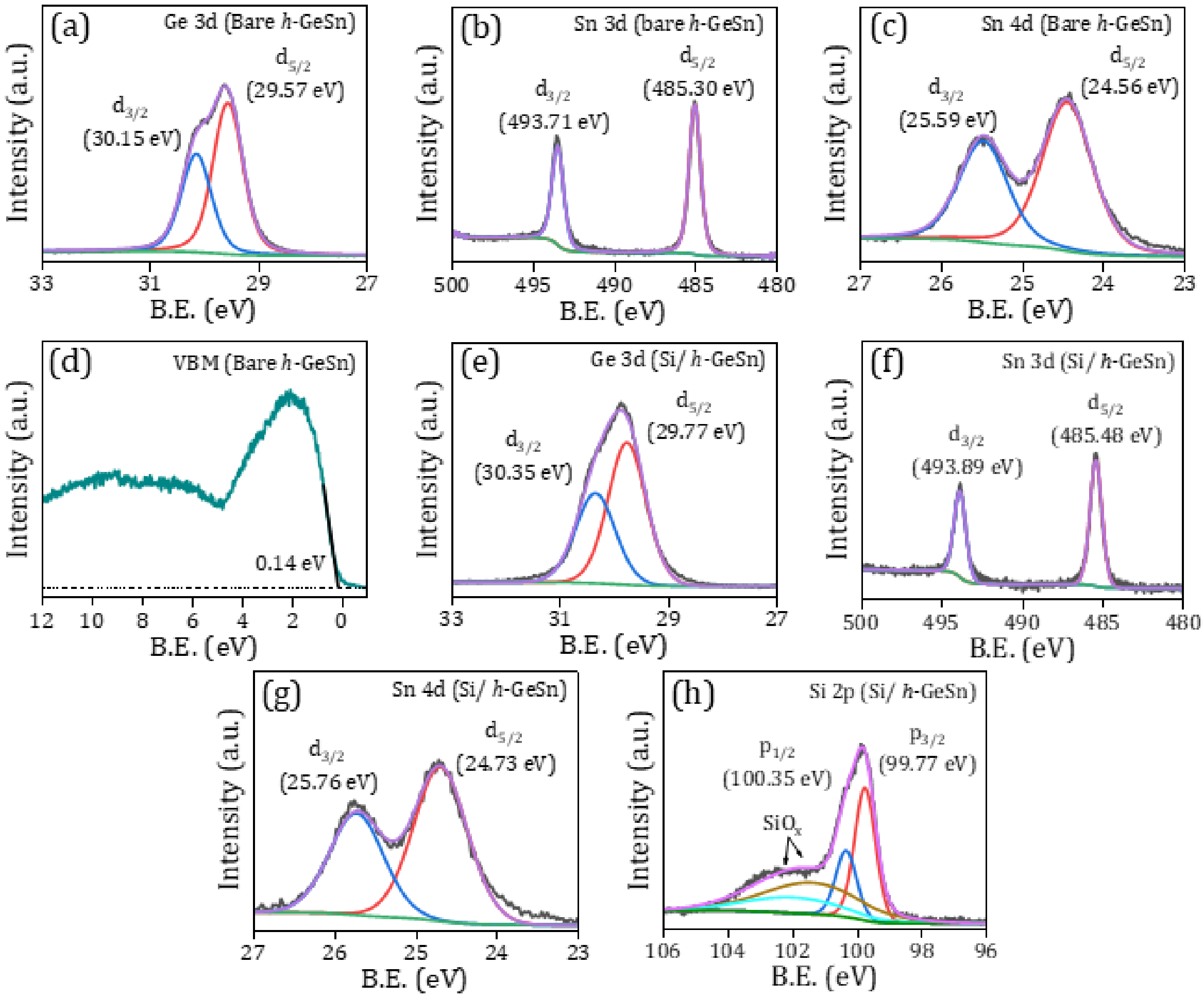

***Figure 6.*** *High-resolution XPS spectra of the bare h-GeSn surface showing (a) Ge 3d, (b) Sn 3d, (c) Sn 4d, and (d) the valence-band maxima. Corresponding core-level spectra acquired at the Si/h-GeSn heterointerface: (e) Ge 3d, (f) Sn 3d, (g) Sn 4d, and (h) Si 2p.*

The extracted core-level positions and VBM values are summarised in Table 1. A shift of the Ge and Sn core levels toward higher binding energy is observed with increasing Sn%, arising from modifications to the local bonding environment and the enhanced lattice distortion introduced by the larger Sn atoms, which reduces core-hole screening and alters the crystal potential. In addition, Sn incorporation redistributes the valence-band density of states, leading to a downward shift of the VBM.

Moreover, the extracted core-level and VBM values were used for the calculation. Valence-band offsets (VBO) for all the fabricated heterostructures were calculated using the Kraut method,[25]

$$\Delta E_V = (E_{CL} - E_{VBM})_{Ge} - (E_{CL} - E_{VBM})_{Si-NM} + (E_{CL} - E_{CL})_{Interface}$$

where $E_{CL}$ and $E_{VBM}$ represent the core levels of materials and the valence-band maximum, respectively.

The extracted VBO values increase from ~0.29 ± 0.02 eV for Si/*l*-GeSn to ~0.39 ± 0.02 eV for Si/*m*-GeSn, and ~0.57 ± 0.02 eV for Si/*h*-GeSn, demonstrating a clear Sn composition dependence of the band alignment. The monotonic increase in VBO with Sn incorporation reflects the observed evolution of the GeSn valence-band electronic structure. The substitution of Ge by Sn introduces Sn $5s^2 5p^2$ orbitals with higher atomic energy and a larger spatial extent than the Ge 4p states, which enhances p-p orbital coupling and shifts the valence-band edge upward relative to the vacuum level.[26] In addition, the larger lattice constant of Sn alters the local bonding configuration and reduces bond strength, thereby decreasing the bonding splitting of the valence states. The combined effects of enhanced p-p orbital interaction, reduced bonding splitting, and lattice expansion collectively shift the GeSn valence-band edge toward higher energy, leading to the observed increase in the VBO with increasing Sn content. Strain also plays an important role; compressive strain in GeSn lifts the heavy-hole/light-hole degeneracy and raises the heavy-hole band energy, further reducing the VBM. This strain-induced valence-band modulation adds to the chemical (alloying) contribution from Sn. As the Si valence-band edge remains essentially unchanged, the upward shift of the GeSn VBM directly translates into a larger VBO at the Si/GeSn interface.

***Table 1.*** *Summary of the Ge 3d, Sn 3d, Sn 4d, and Si 2p core-level binding energies together with the VBM for the bare l-GeSn, m-GeSn, and h-GeSn layers, and for the grafted Si/GeSn heterostructures.*

| Core-level / Material | | VBM (eV) | Ge $3d_{5/2}$ (eV) | Sn $3d_{5/2}$ (eV) | Sn $4d_{5/2}$ (eV) | Si $2p_{3/2}$ (eV) | VBO (eV) |
|---|---|---|---|---|---|---|---|
| Bare Si | | 0.61 | - | - | - | 99.49 | - |
| l-GeSn | Bare | 0.25 | 29.31 | 485.09 | 24.29 | - | - |
| | Heterostructure | - | 29.68 | 485.45 | 24.68 | 99.79 | |
| | | | × | | | × | 0.29 |
| | | | | × | | × | 0.30 |
| | | | | | × | × | 0.27 |
| m-GeSn | Bare | 0.22 | 29.41 | 485.19 | 24.38 | - | - |
| | Heterostructure | - | 29.70 | 485.46 | 24.69 | 99.78 | |
| | | | × | | | × | 0.39 |
| | | | | × | | × | 0.41 |
| | | | | | × | × | 0.37 |
| h-GeSn | Bare | 0.14 | 29.57 | 485.30 | 24.56 | - | - |
| | Heterostructure | - | 29.73 | 485.48 | 24.77 | 99.77 | |
| | | | × | | | × | 0.59 |
| | | | | × | | × | 0.57 |
| | | | | | × | × | 0.54 |

Concluding from the above calculation, the systematic increase in the valence-band offset from 0.29 to 0.57 eV as the Sn composition increases from 8.8% to 12.4%, as summarised in Fig. 7(a). It is noted that the grafted heterostructure contains an ultrathin $Al_2O_3$ bonding layer between the Si and GeSn. Consequently, the extracted band offsets correspond to the effective electronic alignment across the Si/$Al_2O_3$/GeSn heterointerface rather than a Si/GeSn heterostructure. Owing to its ultrathin thickness, the $Al_2O_3$ layer primarily functions as a bonding and passivation layer while maintaining electronic coupling across the heterointerface. PL measurements were performed to obtain insight into the GeSn bandgap; required for the band-alignment analysis (Fig. S7, SI). The PL spectra exhibit the characteristic redshift with increasing Sn content, confirming the progressive reduction in the bandgap and aligning with recent reports.[27-29] The experimentally determined bandgap values were combined with the extracted VBOs to compute the CBOs.[30] As shown in Fig. 7(a), the CBO decreases systematically with increasing Sn composition. It should be noted that Anderson's model assumes an ideal heterointerface referenced to the vacuum level and therefore does not explicitly account for interface dipoles, interfacial chemical bonding, localized band bending, or fixed charges associated with the ultrathin $Al_2O_3$ interlayer. Consequently, the calculated CBOs should be regarded as effective band offsets that provide a first-order description of the electronic structure of the grafted heterointerface. For clearer visualization of the band structure, the corresponding energy-band diagrams were constructed for the grafted Si/*l*-GeSn, Si/*m*-GeSn, and Si/*h*-GeSn heterostructures, as shown in Fig. 7(b-d).

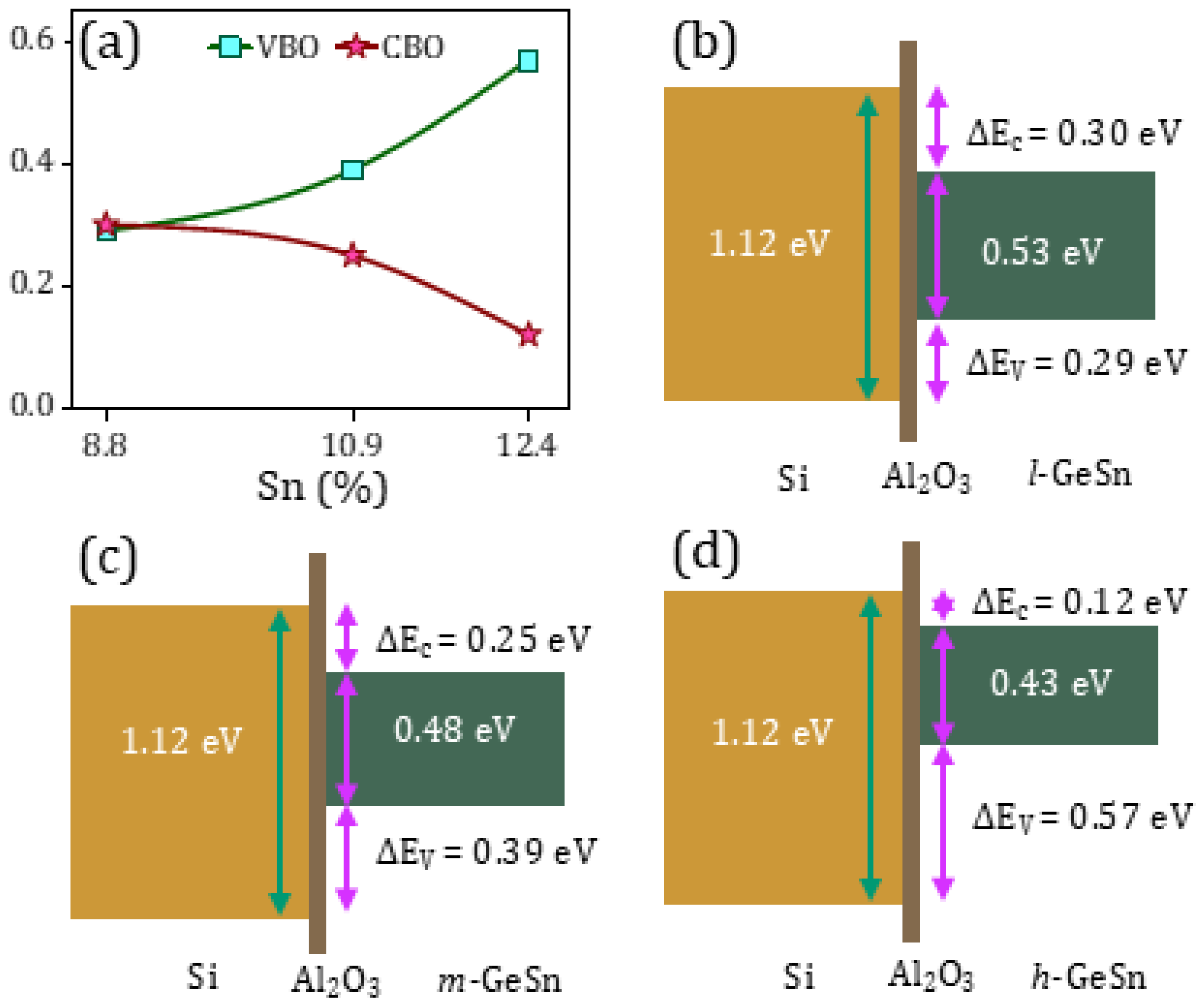

***Figure 7.*** *(a) Extracted Eg, VBO, and CBO as a function of Sn content. Band diagrams for grafted heterostructures Si/GeSn based on (b) 8.8 %, (c) 10.9 %, (d) 12.4% Sn, showing increasing VBO and decreasing CBO.*

## Conclusion:

In summary, we systematically established the composition-dependent band alignment in grafted Si/GeSn heterostructures by combining high-quality interface information with X-ray photoelectron spectroscopy and Kraut's core-level analysis. CMP enabled smooth GeSn surfaces while preserving Sn content, enabling a grafted heterointerface suitable for reliable electronic-structure probing. A clear and monotonic increase in the VBO from ~0.29 to ~0.57 eV was observed as the Sn composition increased from 8.8% to 12.4%, which originates from the alloy-induced modification of the GeSn valence-band electronic states together with the contribution from compressive strain. The resulting band-alignment evolution demonstrates enhanced hole confinement in GeSn and a reduced electron transport barrier at the Si interface, highlighting the capability to tailor interfacial carrier-confinement characteristics through Sn-composition engineering continuously. Correspondingly, the CBO decreases with increasing

Sn%. These findings provide experimental insight into the electronic structure of the grafted Si/GeSn heterointerface formed by semiconductor grafting and establish a quantitative framework for band-structure design in lattice-mismatched group-IV systems. The demonstrated tunability of the band offsets is highly relevant to the development of Si-compatible photonic, electronic, and avalanche devices, where the interfacial band configuration critically determines carrier-selective transport and impact-ionization efficiency.

## Supporting Information:

Optical microscopy images of the fabricated grafted Si/GeSn heterostructures; HRXRD 2θ/ω scans, RSM, and layer schematics used to extract Sn composition of the as-grown GeSn thin films. One table lists the polishing parameters, and the other summarizes the structural and morphological details of the as-grown and polished GeSn epilayers. XPS survey and high-resolution core-level spectra of cleaned bare Si and GeSn surfaces; stepwise $Ar^+$ ion-beam depth-profiling procedures for all three grafted heterostructures together with the change of Si 2p and Ge 3d intensities used to identify the interface, also along with the survey scan, interface core-level spectra of oxygen, carbon, and $Al_2O_3$. The Gaussian-fitted photoluminescence spectra.

## Acknowledgement:

This research was sponsored by Air Force Research Lab through Attollo Engineering under the grant number FA237724CB032.

## Conflict of Interest:

The authors declare no conflict of interest.

## Data Availability Statement

The data that support the findings of this study are available from the corresponding author upon reasonable request.

# Supporting Information

# Composition Driven Band Alignment Evolution in Grafted Si/GeSn Heterostructures

Pargam Vashishtha[1, #], Samuel A. Haessly[1, #], Yang Liu[1], Tsung-Han Tsai[1], Kushal Dahal[2], Shui-Qing Yu[2], Edward, K. Huang[3], Nanditha M. Dissanayake[3], Zhenqiang Ma[1,*]

[1]*Department of Electrical and Computer Engineering, University of Wisconsin-Madison, Madison, Wisconsin 53706, USA.*

[2]*Department of Electrical Engineering, University of Arkansas, Fayetteville, Arkansas 72701, USA*

[3]*Attollo Engineering, Camarillo, CA 93012, USA.*

[#]*Authors have contributed equally*

**Corresponding Author's Email: mazq@engr.wisc.edu*

Fig. S1 showcases the optical microscopy images of the fabricated grafted heterostructures Si/*l*-GeSn, Si/*m*-GeSn, and Si/*h*-GeSn depicted in Fig. S1(a), Fig. S1(b), and Fig. S1(c), respectively. The transferred nanomembranes (NMs) have lateral dimensions of 3.8 × 3.8 mm$^2$.

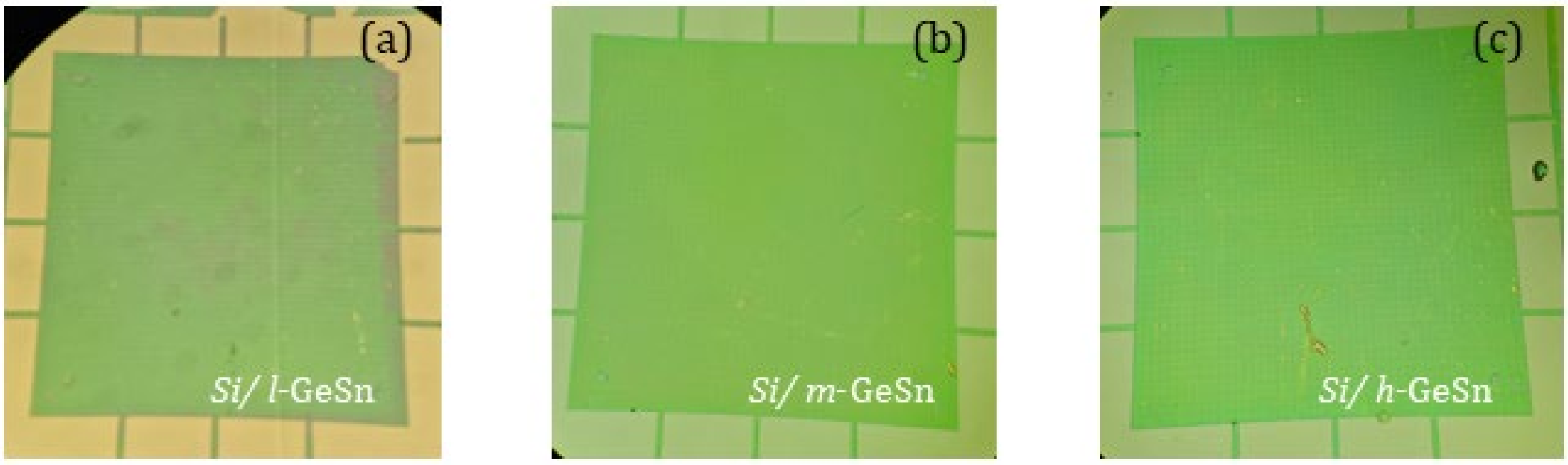

***Figure S1.*** *Optical microscopy images of the fabricated grafted heterostructures: (a) Si/l-GeSn, (b) Si/m-GeSn, and (c) Si/h-GeSn.*

Fig. S2 summarises the structural characterization of as-grown GeSn epilayers with different Sn compositions using HRXRD and RSM, along with corresponding layer schematics. The 2θ/ω diffraction patterns in Fig. S2 (a-c) exhibit the GeSn peaks located on the lower-angle side of the Ge peak, confirming the successful incorporation of Sn into the Ge lattice; the progressive shift of the GeSn peak toward smaller diffraction angles from *l*-GeSn to *h*-GeSn indicates an increase in Sn content and the associated lattice expansion. The RSMs around the asymmetric reflection, shown in panels (d-f), provide direct insight into the strain state and Sn % behavior of the layers, as compiled in Table S2. The Ge virtual substrate lies on the nearly relaxed line (R = 0), while the Si substrate appears as substrate material. The GeSn reflections are located between these two limits, revealing partial strain relaxation that becomes more pronounced with increasing Sn composition, as evidenced by the lateral shift of the GeSn intensity toward larger in-plane lattice parameters. The layer-structure schematics in panels (g-i) illustrate the corresponding heterostructure stacks and Sn compositional distribution, showing the as-grown *l*-GeSn, *m*-GeSn, and *h*-GeSn samples in Fig. S2(g), Fig. S2(h), and Fig. S2(i), respectively.

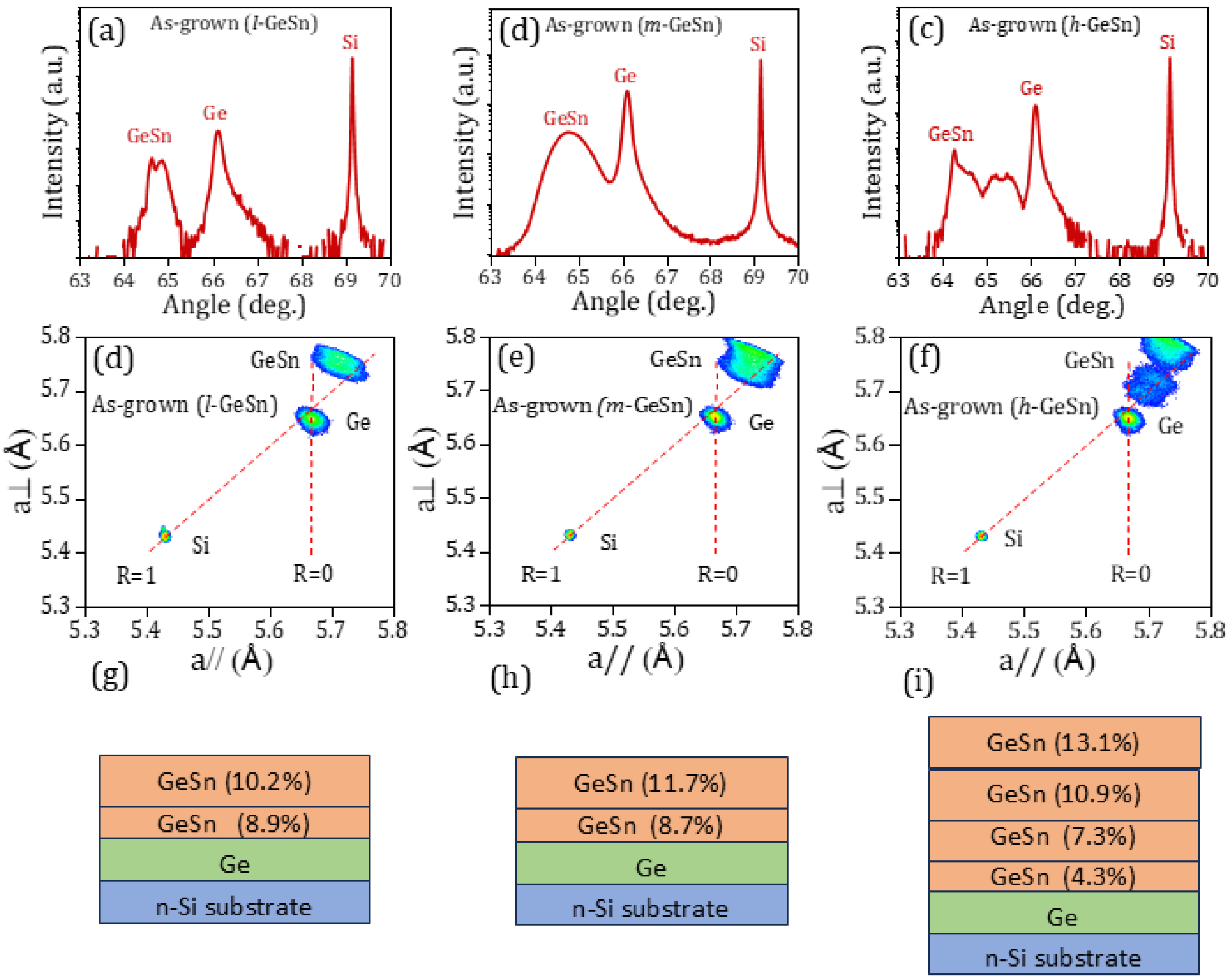

***Figure S2.*** *(a-c) HRXRD 2θ/ω scans of as-grown l-, m-, and h-GeSn; (d-f) Corresponding RSMs indicating progressive strain relaxation of GeSn. (g-i) Schematic of all as-grown GeSn/Ge/n-Si layer structures with the respective calculated Sn %.*

The polishing parameters are summarized in Table S1. The *l*-GeSn and *m*-GeSn samples were polished in a single step for 60 Sec, whereas the h-GeSn sample was polished in two consecutive steps of 15 Sec each. The rotation speed and pre-wet conditions were kept constant for all samples at 30 rpm and 100 mL, respectively. The applied load was 482 gm for *l*-GeSn & *m*-GeSn, and 482 gm for *h*-GeSn. The slurry volumes were 25 mL for *l*-GeSn and *m*-GeSn, while the *h*-GeSn sample received 15 mL in each polishing step. The corresponding etching rates for all the samples were ~0.85 nm/sec.

***Table S1.*** *Compiled roughness, Sn composition, and strain values for the as-grown and polished l-, m-, and h-GeSn epilayers.*

| ***Sample*** | ***Speed*** **(rpm)** | ***Time (sec)*** | ***Weight (gm)*** | ***Slurry (mL)*** | ***Pre-wet (mL)*** |
|---|---|---|---|---|---|
| ***l-GeSn*** | *30* | *60* | *482* | *25* | *100* |
| ***m-GeSn*** | *30* | *60* | *488* | *25* | *50* |
| ***h-GeSn*** | *30* | *15+15* | *482* | *15+15* | *100* |

Table 2 summarizes the surface roughness, Sn composition, and strain state of each sample in both the as-grown and polished conditions. The Sn content and corresponding strain values for all layers from the top GeSn epilayer to the underlying Ge buffer layer are listed, where 0 at.% Sn denotes the Ge buffer. The GeSn layers exhibit compressive strain, as indicated by the negative sign, whereas the Ge buffer layer shows tensile strain.

***Table S2.*** *Compiled roughness, Sn composition, and strain values for the as-grown and polished l-, m-, and h-GeSn epilayers.*

| ***Material*** | | ***Roughness (nm)*** | ***Sn (%) top-to-/ buffer*** | ***Strain (%) top-to-/ buffer*** |
|---|---|---|---|---|
| *As-grown* | *l-GeSn* | *3.93* | *10.2/ 8.9/ 0.0* | *-0.49/ -0.39/ 0.16* |
| | *m-GeSn* | *9.24* | *11.7/ 8.7/ 0.0* | *-0.45/ -0.27/ 01.3* |
| | *h-GeSn* | *7.87* | *13.1/ 10.9/ 7.3/ 4.3/ 0.0* | *-0.63/ -0.45/ -0.15/ -0.08/ 0.16* |
| *Polished* | *l-GeSn* | *0.29* | *8.8/ 0.0* | *-0.34/ 0.15* |
| | *m-GeSn* | *0.31* | *10.9/ 8.4/ 0.0* | *-0.26/ -0.21/ 0.17* |
| | *h-GeSn* | *1.56* | *12.4/ 11.0/ 7.3/ 4.9/ 0.0* | *-0.91/ -0.78/ -0.35/ -0.17/ 0.17* |

Fig. S3 presents the XPS survey and high-resolution spectra of the cleaned bare Si and bare *l*-GeSn surfaces. The survey spectrum of bare Si (Fig. S3(a)) shows only the characteristic Si core-level features without any detectable impurity-related peaks, while the corresponding O 1s and C 1s spectra in Fig. S3(b,c) exhibit no discernible signals. Similarly, the survey scan of bare *l*-GeSn (Fig. S3(d)) reveals the Ge and Sn core-level peaks with no additional contamination-related features. The absence of distinct O 1s and C 1s peaks in Fig. S3(e,f) further confirms a clean and oxide-free GeSn surface.

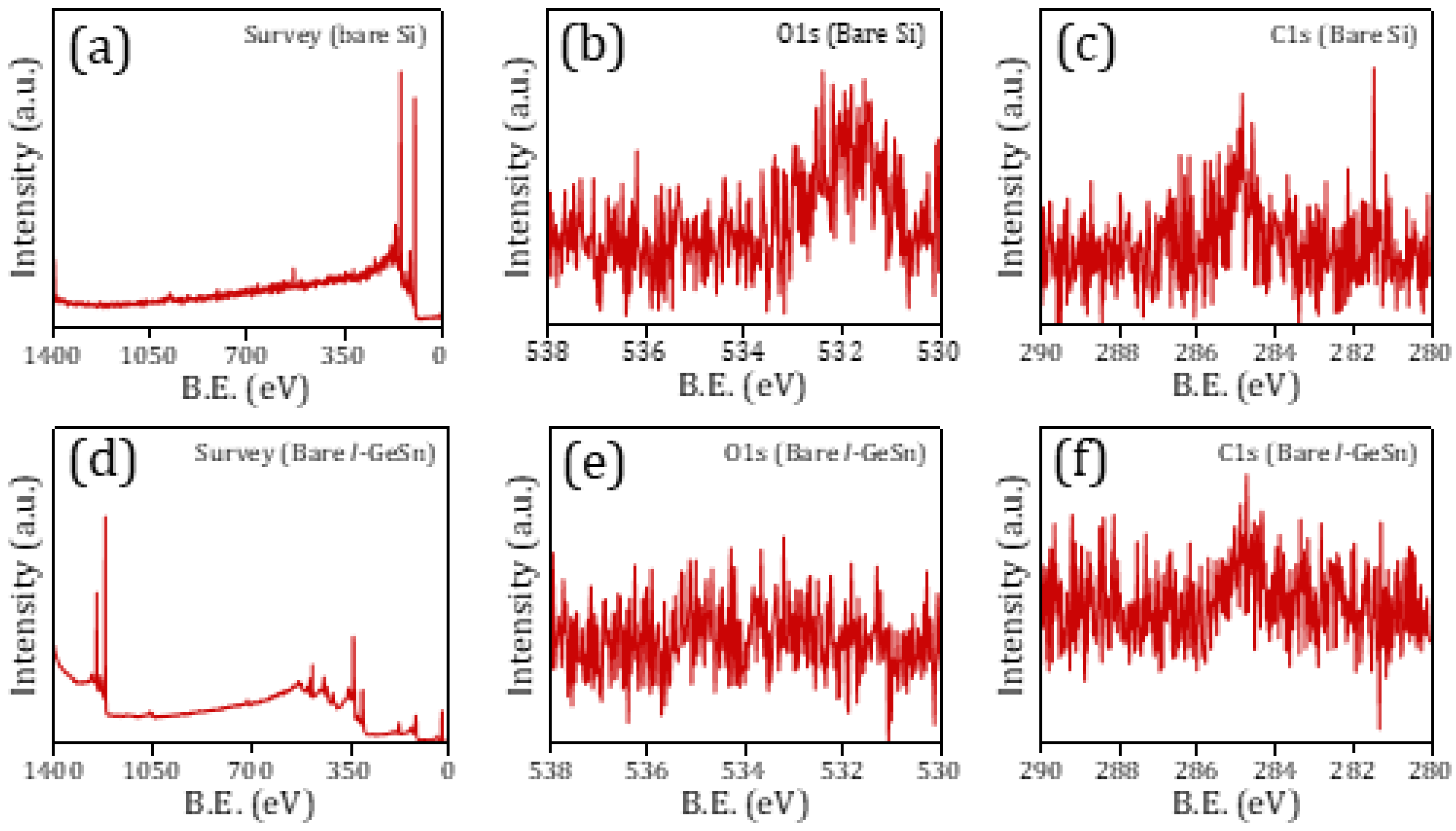

***Figure S3.*** *(a) XPS survey and high-resolution (b) O 1s and (c) C 1s spectra of cleaned bare Si and bare l-GeSn thin film sample; (d) Survey, (e) O 1s and (f) C 1s.*

Fig. S4(a-d) presents the XPS depth-profiling analysis of the grafted Si/*l*-GeSn heterostructure. Fig. S4(a) shows the steps of $Ar^+$ ion-beam energy and duration during the etching process, and the core-level spectra were recorded at each step. The Ge 3d and Si 2p core-level spectra with etching time in panels Fig. 4(b) and Fig. 4(c) reveal a gradual decrease in the Si signal accompanied by a simultaneous increase in the Ge signal, indicating the successful thinning of the Si-NM and the approach toward the grafted Si/GeSn interface. This transition is quantitatively summarised in Fig. S4(d), confirming depth-resolved probing of the heterointerface. After obtaining the interface, the survey spectrum in Fig. 4(e) shows elemental features, with no extraneous peaks. The high-resolution O 1s, C 1s, and Al 2p spectra in Fig. S4(f-h) show no detectable signals, evidencing the absence of oxide and carbon contamination, thereby confirming a clean, oxide-free grafted Si/*l*-GeSn heterostructure.

Fig. S5 presents the XPS analysis of the bare *m*-GeSn surface and the Si/*m*-GeSn heterostructure after depth profiling. Fig. S5(a-c) shows the survey scan and high-resolution C 1s and O 1s core-level spectra of cleaned bare *m*-GeSn, confirming the oxide-free and contamination-free starting surface used for heterostructure fabrication. For the interface measurement, the Si-NM was thinned by step-wise (2k/4k) $Ar^+$ ion-beam etching profile (Fig. S5(d)), and core-level spectra were recorded at each step for Ge3d in Fig. S5(e), and Si 2p in Fig. S5(f) The gradual reduction of the Si 2p signal, accompanied by a simultaneous increase in the Ge 3d intensity as complile in Fig. S5(g), reveals a controlled approach toward the Si/GeSn interface and enables depth-resolved probing of the heterostructure. The corresponding intensity variation quantitatively confirms the interface position. The survey spectrum (Fig. S5(h)) collected at the interface shows elemental features, while the high-resolution O 1s (Fig. S5(i)) and C 1s (Fig. S5(j)) spectra exhibit no detectable signal, evidencing the absence of oxide and carbon contamination and thereby confirming a clean Si/*m*-GeSn heterointerface.

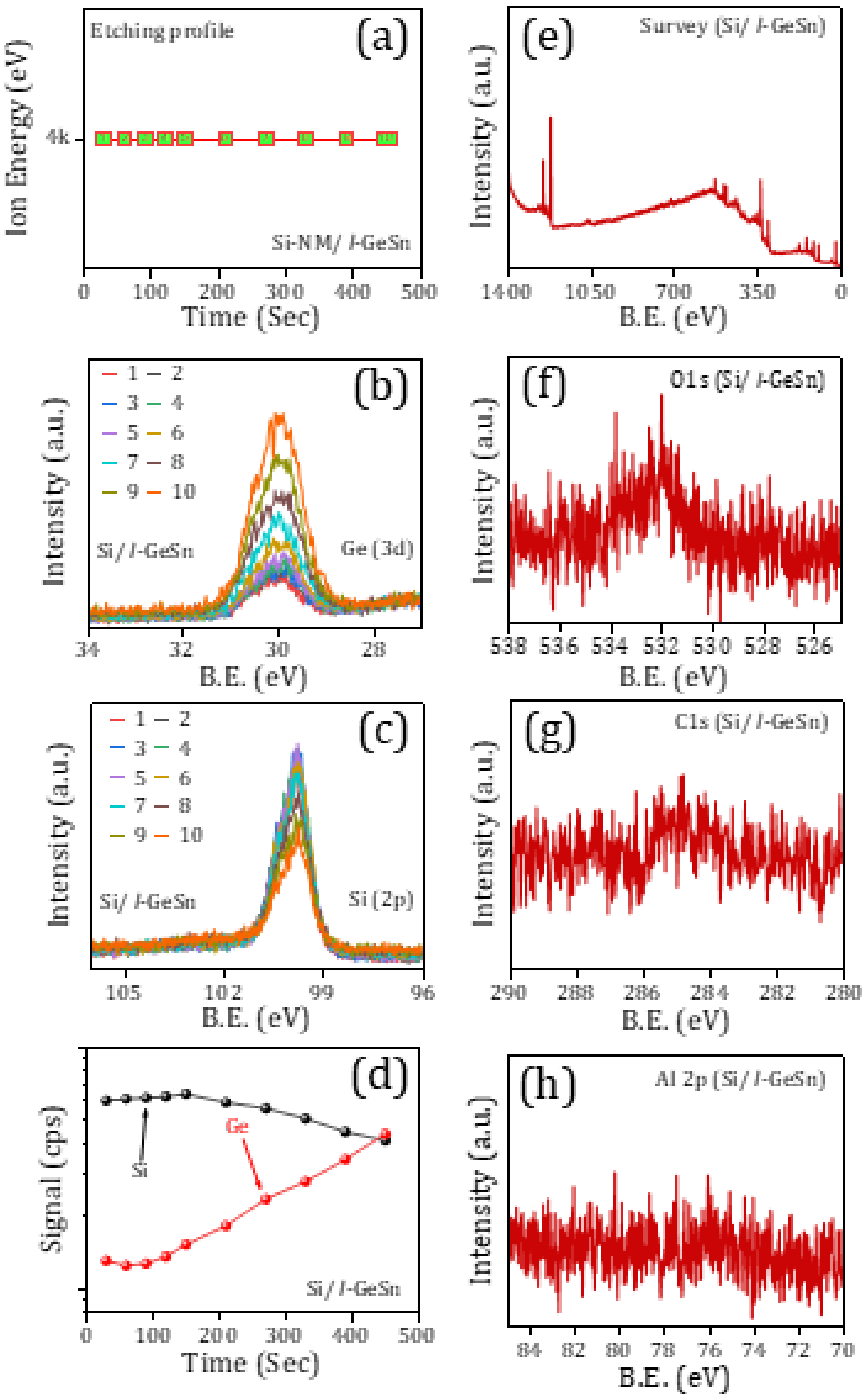

***Figure S4.*** *(a) Step-wise ion-beam energy with respect to time, (b) Ge 3d and (c) Si 2p core-level spectra at each etching step. (d) Corresponding Si and Ge signal intensity variation with etching time. (e) Survey (f) O 1s, (g) C 1s, and (h) Al 2p spectra at the interface of the grafted Si/l-GeSn heterostructure.*

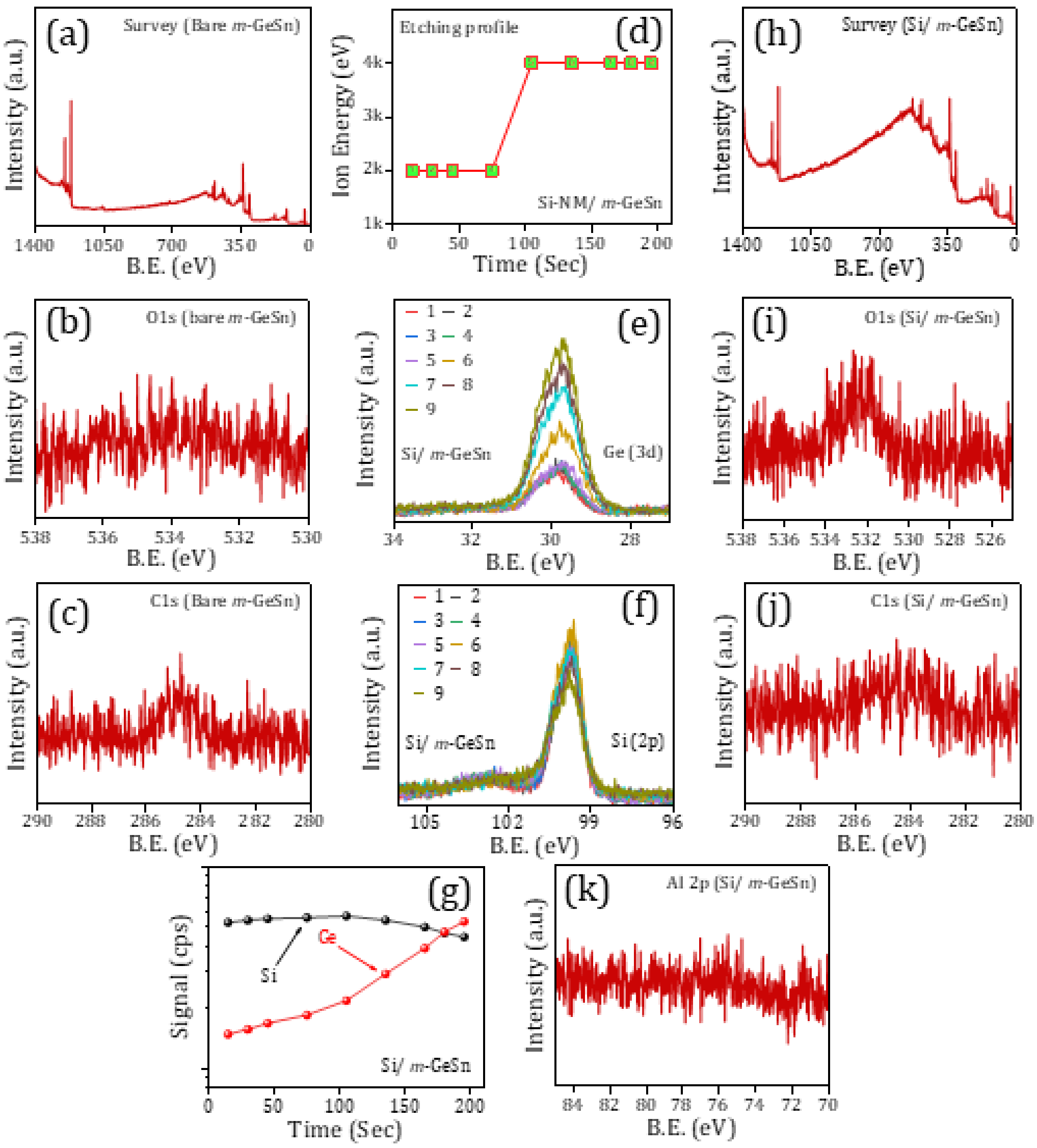

***Figure S5.*** *(a) Survey, and (b) O 1s, (c) C 1s core-level spectra of cleaned bare m-GeSn. XPS (d) depth profiling, showing the change of (e) Ge 3d and (f) Si 2p with etching step, (g) etching time-corresponding signal intensity variation for the grafted Si/m-GeSn heterostructure, and (h) survey, (i) O 1s, (j) C 1s, (k) Al 2p spectra at the interface.*

Fig. S6 shows the XPS investigation of the bare *h*-GeSn surface and the Si/ *h*-GeSn heterostructure following depth profiling. The survey spectrum together with the high-resolution C 1s and O 1s spectra of the in situ cleaned bare *h*-GeSn (Fig. S6(a–c)) confirm a contamination-free and oxide-free starting surface before heterostructure formation. To probe the interface, the Si-NM was thinned using a step-wise 4 kV $Ar^+$ ion-beam etching sequence, as illustrated in Fig. S6(d), and core-level spectra were collected after each etching step. The change in the Ge 3d and Si 2p signals (Fig. S6(e,f)) shows a continuous decrease in the Si 2p intensity accompanied by a corresponding increase in the Ge 3d intensity. The compiled intensity variation in Fig. S6(g) clearly indicates the controlled approach toward the interface and enables depth-resolved analysis of the heterostructure. The survey spectrum acquired at the interface (Fig. S6(h)) displays only the expected elemental features. In addition, the high-resolution O 1s and C 1s spectra (Fig. S6(i,j)) show no detectable peaks, confirming the absence of native oxide and carbon.

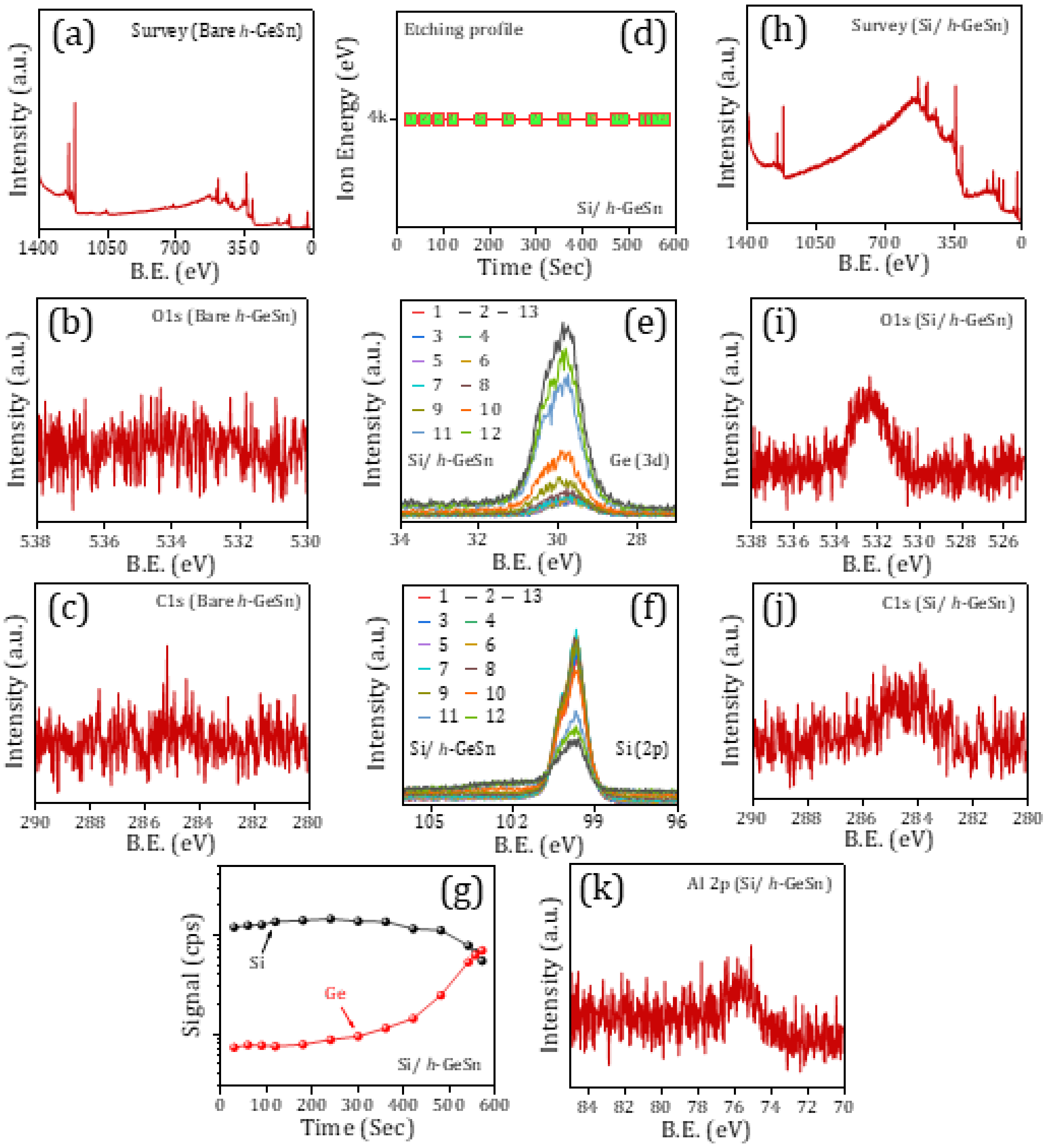

***Figure S6.*** *(a) Survey, and (b) O 1s, (c) C 1s core-level spectra of cleaned bare m-GeSn. XPS (d) depth profiling, showing the change of (e) Ge 3d and (f) Si 2p with etching step, (g) etching time-corresponding signal intensity variation for the grafted Si/m-GeSn heterostructure, and (h) survey, (i) O 1s, (j) C 1s, (k) Al 2p spectra at the interface.*

The room-temperature (300 K) PL spectra of the GeSn epilayers exhibit an optical bandgap that increases with increasing Sn incorporation, as shown in Figure S7 (Gaussian-fitted data). For the *l*-GeSn sample (8.8%), the band gap is taken to be 0.53 eV from the recent literature.[1-5] In contrast, the PL spectra recorded for the *m*-GeSn (See Figure 7(a)) were deconvoluted into 2 peaks, where 0.48 corresponds to the band gap, while the other is the lower Sn % layer. The strong emission peak at ~0.48 eV confirms enhanced optical quality and a transition toward direct-gap-dominated recombination. For the *h*-GeSn sample, the PL deconvolved spectra (showcased in Figure S7(b)), along with the band gap emission peak at ~0.43 eV, show other emission peaks from different Sn% layers (~0.46, ~0.50, and ~0.56 eV), which can be attributed to different compositional layers in this sample, as observed in our HRXRD structural examination. The low-energy peak (~0.43 eV), corresponding to the highest Sn-content top layer (~12.4%), shows a weak intensity due to its very thin thickness. Overall, the PL results indicate a systematic redshift (see Figure S7(c)) of the bandgap compared with the low-Sn sample and a significant improvement in radiative recombination for intermediate Sn incorporation, while peak broadening at the highest Sn composition reflects increased alloy disorder and band structure perturbation.

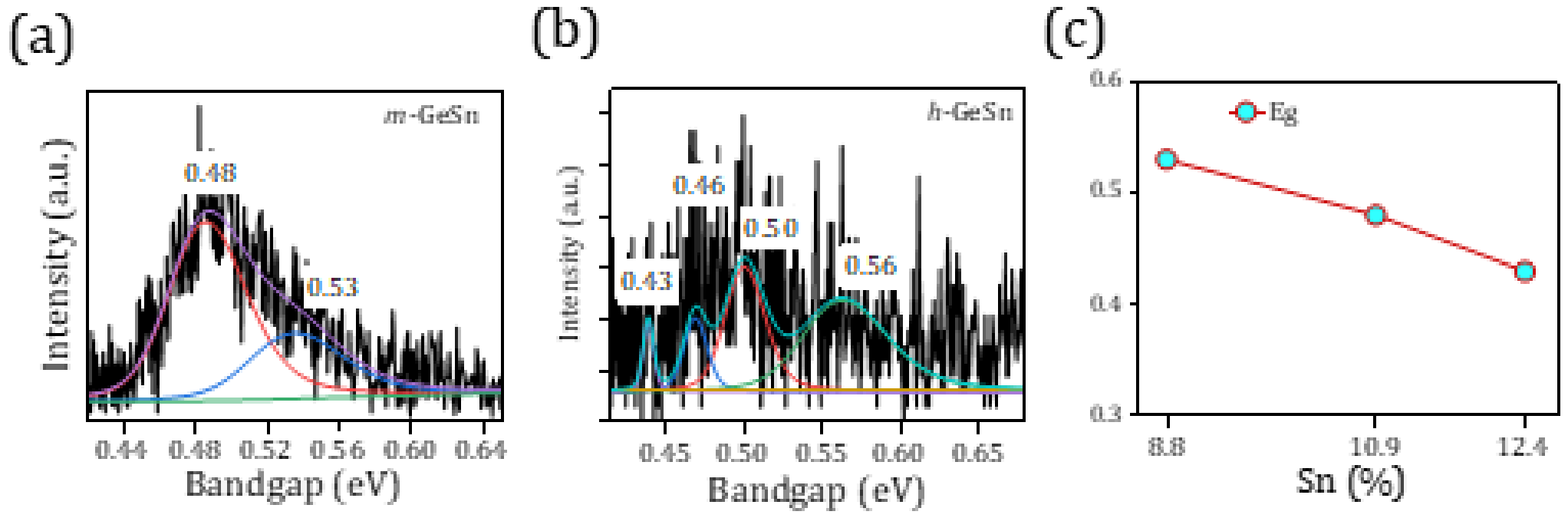

***Figure S7.*** *Gaussian-fitted PL spectra of (a) m-GeSn and (c) band gap vs. Sn % spectra.*

## Refrences